\documentstyle[11pt,paspconf,epsf]{article}
\begin{document}

%personal definitions

\def\myfig #1#2#3#4{\par
\epsfxsize=#1 cm
\moveright #2cm
\vbox{\epsfbox{#3}}
{\noindent Figure~#4 }\vskip .3cm }
%---------------------------
\def\bh{black~hole~}
\def\ad{accretion~disk~}
\def\el {emission-line~}
\def\ip{ionization-parameter~}
\def\ed {Eddington~}
\def\ers{{\rm erg/sec}}
\def\ms{M_{\odot}}
\def\et{{\it et.al.}}
\def\sw{Schwarzshild~}
\def\bb{black body~}
%----------------------
\title{ The Mass-Luminosity Relation in AGN
}
% Review presented during the meeting
%"Structure and Kinematics of Quasar Broad Line Regions"
%23-26 March 1998, Lincoln, Nebraska

\author{A. Wandel}
\affil{ Racah Institute, The Hebrew University, Jerusalem 91904,
Israel}

\begin{abstract}
Probably the most fundamental characteristic of the quasar-AGN power house, the
mass of the central black hole, is the least well known. The broad emission lines have
been, and probably will remain, our best probe of the central mass. Using
these probes to estimate the \bh mass suggests that over more than six orders
of magnitude, the ratio between the continuum luminosity and the central mass
(the Eddington ratio) has a spread  perhaps as narrow as 1--2 orders of
magnitude, while other methods give a larger spread and possibly a
luminosity-dependent \ed ratio. I review the
three main classes of mass estimation methods---BLR
kinematics, X-ray variability and \ad modeling, and their results for
$L/L_{\rm Edd}$. Potential sources of error and biases are discussed.
\end{abstract}


%\section{Introduction}

\section{The $L/M$ relation and mass estimation methods}

The most fundamental, yet the least well known, parameter
of the AGN central engine is its mass.
While the AGN luminosity, which is readily measurable,
varies over six orders of magnitude, the ratio $L/M$
(or, alternatively, the \ed ratio, $L/L_E$) apparently
has a rather small range, indicating that the $L/M$ value
is a property characterizing the AGN phenomenon.
In terms of the \ed ratio we find from various mass
estimation methods and for various AGN samples
$L/L_E\sim$~0.001--1. This value and its spread depend on
the method used to estimate the mass.

The methods used to determine the central mass in AGN
can be divided into three major groups:
BLR kinematics, X-ray variability and \ad spectral fitting.
The first group uses the broad emission lines as probes,
assuming the velocity dispersion of the line-emitting gas
is induced by the gravitational potential of the central mass.
The second gives the variability mass limit (eq. \ref{equ:mdt}),
while the third is a variant of the  temperature mass
(eq. \ref{equ:mt}), essentially trying to fit the UV bump with
a spectrum from a thin \ad model. In the following sections
I will describe the three groups and
the problems associated with each of them.

\section{Some fiducial masses}
Any compact, accretion-powered radiation source can be assigned several fiducial mass
estimates:

1. The \ed limit. In order to maintain steady spherical accretion the luminosity must be less than
the \ed luminosity, $L<L_{\rm Edd}=4\pi GMm_pc/\sigma_T=1.3\times 10^{46}M_8$~erg~s$^{-1}$, or
$M_8>L_{46} $
where $ M_8=M/10^8\ms$ and $L_{46} =L/10^{46}$~erg~s$^{-1}$.

2.  The variability limit. The shortest time scale for global variations in the luminosity is the light
travel time across the Schwarzschild radius, $R_s/c=2GM/c^3=10^{3}M_8$~s. Hence if the
luminosity is observed to vary significantly on a time scale $\Delta t$, the \bh mass has to be
\begin{equation}
\label {equ:mdt}
M_8<(\Delta t/10^3 r^{-1}~{\rm s}),
\end{equation}
where $r$ is the effective radius of emission in units of $R_s$.

3. The temperature mass. If a luminosity $L$ comes from an \ad region of radius $R$ and
temperature $T$, then $L<4\pi R^2\sigma T^4$. If a spectral feature at frequency $\nu$ is due
to emission from a \bb \ad at a characteristic radius $R$, then the temperature is given by
$h\nu=3kT$, and we have an upper limit on the \bh mass:
\begin{equation}
\label {equ:mt}
M_8>100 (T/10^5~{\rm K})^{-2}L_{46}^{1/2} (R/R_s)^{-1}.
\end{equation}

4. The evolutionary mass. This mass involves the evolution of a
massive \bh due to accretion. We may define the
\ed time $t_E$ as the time required for an accretion-powered
object radiating at the \ed luminosity to double its mass,
$
t_E=Mc^2/L_E=4\times 10^8 {\rm y}.
$
The observed luminosity implies an accretion rate of
$
\dot M = 0.16 \epsilon^{-1} L_{46} \ms {\rm y}^{-1}
$
where  $\epsilon$ is the efficiency.
Combining these two expressions gives the evolutionary mass,
accumulated if the accretion rate is maintained during a time $t_E$ :
\begin{equation}
\label {equ:mev}
M_{8} = 0.6 \epsilon^{-1} \eta^{-1} f_L L_{46}
\end{equation}
where the parameters $\eta$ and $f_L$  are the \ed luminosity ratio and the fraction of the active
time for intermittently active AGN.

\section{ BLR kinematics}

The broad emission lines characteristic of quasar and
Seyfert~1 spectra are believed to originate from gas highly ionized by the
continuum radiation of the central source. The line width must be due to bulk
motions, probably an ensemble of clouds
moving in the gravitational potential of the central \bh .
Alternative scenarios include radiative acceleration (e.g.
Blumenthal and Mathews 1975), which would induce radial outflow.
However, radial motions are probably excluded because they would induce
asymmetric line profiles (if the emitting clouds are optically thick) which are
usually not observed.
Most workers today agree that the line width in AGN is induced by Keplerian
bulk motions. If this is the case, knowledge of the distance of the
line-emitting gas from the central mass would allow us to estimate the \bh mass,
\begin{equation}
\label {equ:mg}
M \approx v^2R/G .
\end {equation}
The kinematic mass-estimation methods can be divided into two sub-groups,
depending on the method used to find the \el distance:
photoionization methods and reverberation mapping.

\subsection{Photoionization: the \ip method }
We assume that the spatial extent of the BLR
may be represented by a characteristic size --- e.g. the radius at which
the emission peaks.
Indeed, reverberation analyses have shown that
the time lag of emission-line variability varies depending on the line,
and the emission region may be extended in radius.
However, photo-ionization calculations have shown that the
line emissivities of many lines peak sharply in a relatively
narrow range of density and ionization parameter (Baldwin et~al. 1995).
Depending on the radial variation of the density this may imply
that the emission of a given line is localized in a relatively narrow shell
around the continuum source.  Given these results, the single radius model may be
a reasonable approximation for a particular emission line.
The physical conditions in the line-emitting gas are largely determined by
the ionization parameter (the ratio of ionizing photon
density to the electron density, $n_e$, e.g. Netzer 1990)
$
U = Q_{ion} /{4\pi r^2} c n_e
$
where the ionizing photon flux (number of ionizing photons per unit time) is
$
Q_{ion}= \int_{E_0}^\infty F(E) {dE \over E}.
$
$F(E)$ is the luminosity per unit energy,
and $E_0$ is the ionization potential of the line
under consideration (1 Rydberg=13.6~eV for the hydrogen lines).
Defining the ionizing luminosity,
$L_{ion} = \int_{E_0}^\infty F(E) dE$
and the mean energy of an ionizing photon,
$\bar E_{ion} \equiv L_{ion}/Q_{ion}$,
the BLR radius may be written as
\begin{equation}
\label {equ:rion}
r= \left( {L_{ion}\over 4\pi c\bar E_{ion} U n_e}\right ) ^{1/2}
= 13.4  \left ( {L_{44}\over U n_{10}  \epsilon} \right )^
{1/2} ~~{\rm light-days}
\end{equation}
where $n_{10}=n_e/10^{10}$~cm$^{-3}$, $L_{44}=L_{ion}/10^{44}$~erg~s$^{-1}$,
and $ \epsilon =\bar E_{ion}/E_0$ is the mean photon energy in units
of $E_0$.
Combining equations \ref{equ:mg} and \ref{equ:rion} we get the expression for
the \bh mass (in units of $10^8\ms)$:
\begin{equation}
\label {equ:mion}
M_8= 0.5 v_3^2  \left ( {L_{44}\over U n_{10}  \epsilon} \right )^{1/2}
\end{equation}
where $v_3=v(FWHM)/10^3$~km~s$^{-1}$.
Analyses of the broad emission lines in various AGN indicate that
typical values in the gas emitting the high excitation broad lines are
$U\sim$~0.1--1 and
$n_e \sim$~$10^{10}$--$10^{11}$~cm$^{-3}$ (e.g. Rees, Netzer \& Ferland, 1989),
so that $Un_{10}\sim$~0.1--10.
For a given ionizing spectrum, one can explicitly calculate
the photon flux $Q_{ion}$, or alternatively
$ \epsilon$ and $L_{ion}$ in eq. \ref{equ:rion}.
The ionizing flux is dominated by the EUV continuum in the 1--10
Rydberg regime, where most of the ionizing photons are
emitted. Since the continuum
in this range cannot be observed directly, it may be estimated by
extrapolation from the nearest observable energy bands:
the UV and soft X-rays.
The far UV spectrum has been observed beyond
the Lyman limit for about 100 quasars
to wavelengths of 600~\AA\ (Zheng et~al. 1996).
Wandel (1997) uses an estimate of the ionizing flux
based on the soft X-ray continuum spectrum. This approach
is particularly appropriate for lines with high ionization
potentials such as C~{\sc iv} and O~{\sc vi}, which are
derived from more energetic photons,
thus closer to the soft X-ray band.

\subsection{The emission-volume method }

A different approach that does not require the thin shell assumption tries to
estimate the size of the emitting volume. Assuming this volume to be roughly
spherical, the line luminosity would be (Dibai 1981)
\begin{equation}
\label {equ:lvol}
L_{line} = {4\pi\over 3} R^3 f_v j_{line} ,
\end{equation}
where $ j_{line}$ is the line volume emissivity and $f_v$ is the volume filling
factor. Dibai (1981) took $f_v=0.001$ which is arbitrary and probably an overestimate.
Wandel \& Yahil (1985) elaborated the method by expressing $f_v$ in terms of
the better known $f_a$, the angular covering factor:
$
f_v \approx\frac {f_a N}{nR} ,
$
where $N$ is the column density of the line emitting gas.
Substituting this into equation \ref{equ:lvol} gives (for the H$\beta$ line)
\begin{equation}
\label {equ:rvol}
R = 15 \left (\frac {L(H\beta)} {10^{42}~{\rm erg~s^{-1}}}\right ) ^{1/2}
(n_{10} N_{23} {f_a\over 0.1}) ^{-1/2} ~{\rm light-days},
\end{equation}
where $N_{23}=N/10^{23}$~cm$^{-2}$.
The implied kinematic \bh mass is obtained by combining this with eq.~\ref{equ:mg},
\begin{equation}
\label {equ:mvol}
M_8= 0.4 v_3^2  \left ( {L_{42}(H\beta)\over
n_{10} N_{23} {f_a/ 0.1}}\right )^{1/2} .
\end{equation}
Applying this method for a sample
of about 90 low-redshift AGN (mainly Seyfert~1
galaxies and quasars), Wandel \& Yahil (1985) found a tight correlation between
the \bh mass (inferred from the H$\beta$ line)
and the absolute magnitude (fig. 1).
%(fig~ \ref{wy})).
A similar value may be obtained from the sample of Joly et~al. (1985).
%
%\begin{figure}

\myfig {10}{2}{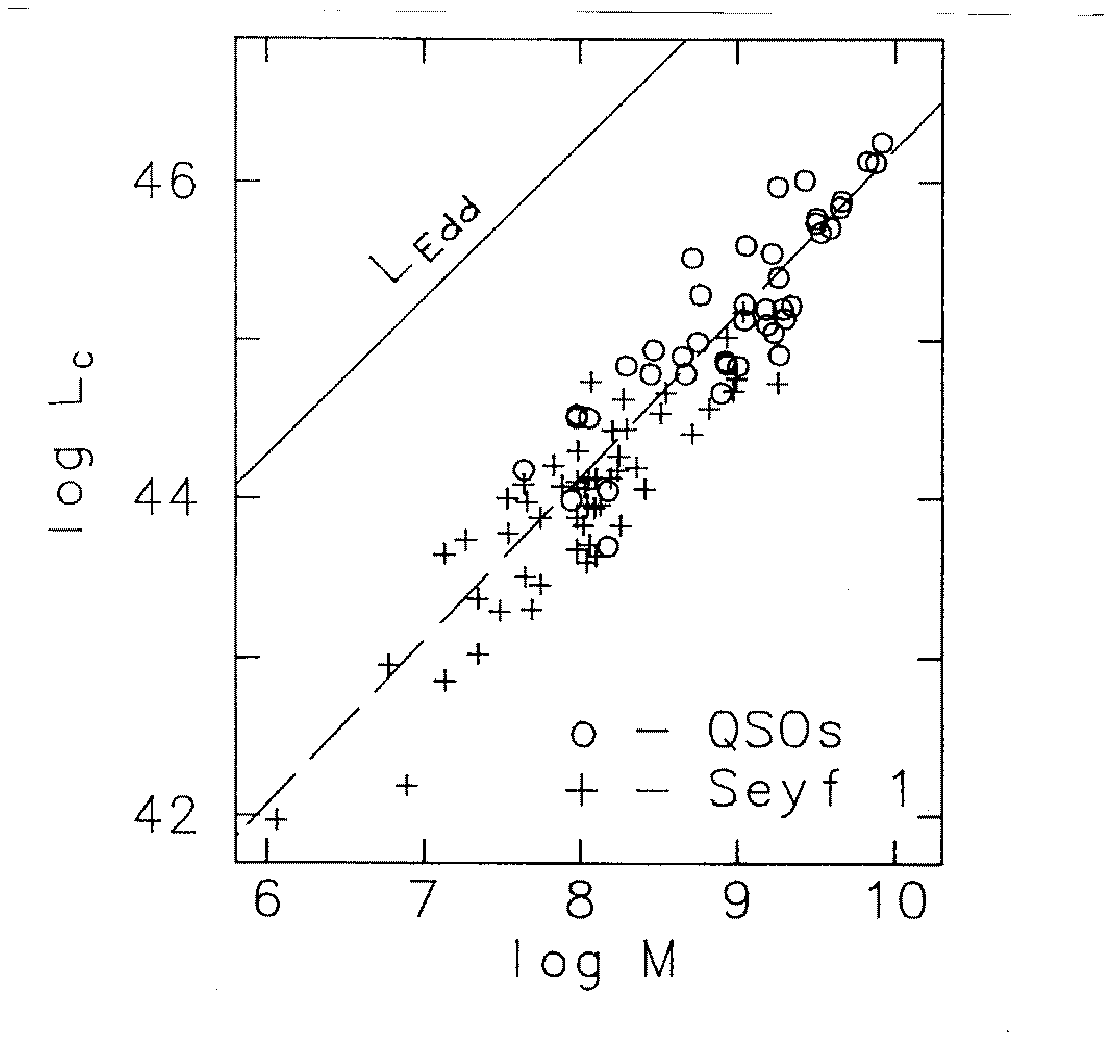}
{1. Optical luminosity versus the mass calculated with the emission volume method
from the H$\beta$ line.}

%\plotone{wy.PS}%\label{wy}
%\end{figure}
Although in part this correlation is
expected, because of the known strong correlation between the line and
continuum luminosities, the correlation found is actually steeper and stronger
than the expected one. If true, this correlation may imply a universal
mass-luminosity relation, which, in terms of the Eddington luminosity
($L_{\rm Edd}=1.3\times 10^{45}M_8$~erg~s$^{-1}$), can be expressed as
$\log L/L_{\rm Edd}= -2\pm 0.5$.

\subsection{The narrow emission lines}

Similar methods can be applied to the narrow emission-line region (NLR),
provided the gas there is gravitationally bound. Typical parameter values for
the NLR are $Un_{10}=10^{-10}$ and $f_a=0.01$ (e.g. Brotherton et~al. 1994).
Of course, within the distance of the NLR from the center, the stellar mass of the
host galaxy could be comparable to the mass of the central black hole, but they may
be correlated (see Laor 1998). Wandel \& Mushotzky (1986) find
an excellent correlation between the mass estimated from the narrow-line width
and the \bh mass estimated from X-ray variability (fig.~3 in section 4 below).
Wandel \& Mushotzky (1986) also apply this method to a sample of Seyfert 2 galaxies
finding that Seyfert~2s as a group
obey the same correlation, provided we observe only a fraction (0.1) of the
emitted X-ray continuum. This is consistent with presence of an obscuring torus,
as suggested by the unified scheme of AGN.

\subsection{ Reverberation mapping }

Variability in the continuum UV luminosity is observed to be followed by
corresponding changes in the line luminosity with delays ranging from days in
low-luminosity AGN to months in luminous quasars. Cross-correlating the continuum
and \el light curves gives the peak of the power in the delay and the
characteristic size ($R_{\tau}=c \tau$) of the \el region. Using this size in
eq. \ref{equ:mg} gives an alternative method to calculate the \bh mass.
Combining all the objects with reliable reverberation data, it appears that
$R\propto L^{1/2}$ (Peterson 1994; Kaspi 1997). With this dependence in mind, eq. \ref{equ:mg}
gives a similar empirical dependence to that in eqs. \ref{equ:mion} and \ref{equ:mvol}:
$M\propto v^2 L^{1/2}$.
The effective BLR radii for the reverberation and \ip methods (both for the H$\beta$ line)
have been found to agree very well, particularly when the shape of the ionizing
continuum is taken into account (Wandel 1997). So do of course the inferred masses.

\subsection{The variable component of the line profile: RMS spectrum}

Although giving a more direct measurement of the BLR size than the
photo-ionization methods, the reverberation mapping kinematic method has two
drawbacks:

\begin{itemize}
\item
Very few AGN (about a dozen) have reliable reverberation data.
Measuring the reverberation radius requires
many observations and is a large project.
\item
It is not clear that the line variability size measured by the
technique of reverberation (or echo) mapping is
the same one contributing the \el profile.
\end{itemize}

The line profile is given by
\begin{equation}
\label {equ:fdlam}
f(\Delta \lambda)= 4\pi\int_{R_{in}}^{R_{out}}\delta [v(R)\sin(i) -
c(\Delta \lambda /\lambda )] E(R)R^2dR
\end{equation}
where $E(R)$ is the volume line emissivity at a radius $R$ and $i$ is the inclination
angle to the line of sight.
This may be used to define  a velocity-weighted radius, e.g. $R_{FWHM}$,
associated with the FWHM of the line profile.
A related radius is the emissivity-weighted radius, that may be defined by
\begin{equation}
\label {equ:rem}
R_{em} = \int_{R_{in}}^{R_{out}}E(R)R^3dR~ / \int_{R_{in}}^{R_{out}}E(R)R^2dR.
\end{equation}
The reverberation radius is obtained by cross correlating the line
and continuum light curves.

For a thick geometry these radii may be quite different. One solution
to this difference is to combine the reverberation-mapping radius with the
velocity inferred from the variable component of the line profile.
In this method, the distance of the line-emitting gas from the cetral source
(the reververation radius) and its velocity dispersion (given by the the rms
spectrum, which reflects only the variable part of the emission-line profile)
both refer to the same gas. Comparing the
rms reverberation  and the ionization-parameter methods, one may
calibrate the latter, which is easier to use and may be applied also for
objects without variability data.

\subsection{Uncertainties and potential error sources}
There are several drawbacks which introduce possible errors into the \el kinematic methods:
\begin{itemize}
\item Uncertainty and scatter in the parameters.
In the ionization method we do not know the values of $n$ and $U$.
\item In the volume-emission method the parameters $N$, $f_a$ and $n$ are uncertain.
\item The velocity of the emitting gas may be anisotropic.
\item  The illumination of the clouds may be anisotropic.
\item The motions may be due to non-gravitational forces.
\end{itemize}

\section{Time variability}

The AGN continuum is known to vary on time scales that depend on the wavelength and
luminosity. While for the UV the time scale ranges from days to months (depending on the
luminosity), the time scale for variations in X-rays is of the order of hours.
As we have seen above, the variability can be used to place a lower limit on the \bh mass. If we
knew the radial dependence of the local X-ray spectrum in the inner disk, the effective emission
radius in eq. \ref{equ:mt} would be
\begin{equation}
\label {equ:rad}
R_{eff} = \int_{R_{in}}^{R_{out}}F_x(R)R^2dR~/ \int_{R_{in}}^{R_{out}}F_x(R)RdR
\end{equation}
where $ F_x(R)$ is the X-ray emissivity per unit area from the disk at radius $R$.
Since there is no widely accepted model for the X-ray continuum emission in AGN, we assume
$R_{eff}$ is a few \sw radii, e.g. 5$R_s$, where the emission from the thin disk around a \sw \bh
peaks. For this value eq.~\ref{equ:mdt} gives $M_8< 0.5\Delta t/10^4$~s.
Barr \& Mushotzky (1986; fig. 2) have found a good
correlation between $\Delta t$ and the X-ray
luminosity.
\vskip 1cm
%\begin{figure}
%\vspace {10cm}\plotone{bm.ps}\caption
\myfig  {10} 1 {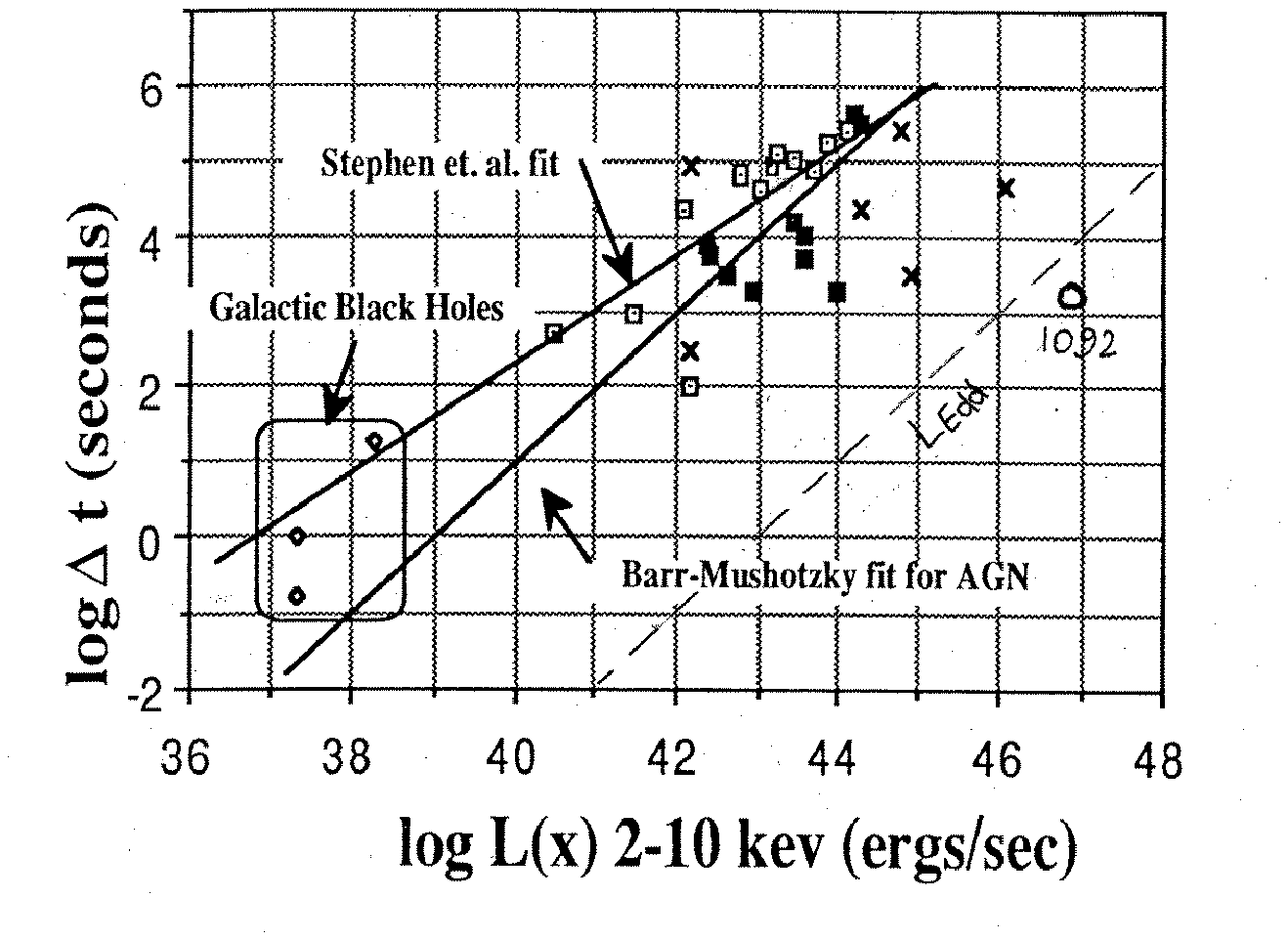}
{2. X-ray variability time scale versus X-ray luminosity. Circles --- Barr \& Mushotzky (1986), squares ---
{\it EXOSAT\/} data, crosses --- radio-bright objects.
Solid lines are best fits (with and without Galactic black holes).
The dashed line indicates the \ed limit assuming $R=5R_s$.}
%\label {bm}\end{figure}
While the variability masses of quasars and Seyfert~1s seem to give Eddington ratios well below
unity, some bright radio objects (e.g. BL Lacs) as well as some
Narrow-Line Seyfert~1 galaxies (NLS1)
have values close to unity and even larger than unity,
e.g. PHL~1092 (see Brandt \& Boller 1998).
Taking into account the data on Galactic black holes and bright radio sources,
one gets a smaller slope in the $L$-$\Delta t$ plane, which may be interpreted
(assuming $M\propto \Delta t$) as the $L/M$ ratio increasing with luminosity.
It may therefore be of interest to compare the masses derived by the
variability method (which admittedly gives
only an upper limit) with the predictions of other methods,
in order to find how tight the upper limits are, or how well they correlate
with the real mass.
Wandel \& Mushotzky (1986) find a very good linear
correlation between the X-ray variability time
and the kinematic mass (calculated from the [O~{\sc iii}] line; see fig. 3),
%\ref{wm}),
which supports the hypothesis that $\Delta t$ (X-ray) and the FWHM
of [O~{\sc iii}] both are closely related to the central mass.
Interestingly, in the sample used by Wandel \& Mushotzky (1986),
the correlation of $\Delta  t$ with the mass derived from H$\beta$ is less
good than that with the mass derived from [O~{\sc iii}].
For broad lines, a good correlation between
$\Delta  t$ (X-ray) and the H$\beta$ line width is obtained
in a mixed sample containing rapidly variable NLS1 and normal Seyfert~1s
(Wandel \& Boller 1998).

%\begin{figure}
%\vspace {9cm}\plotone{wm.ps}\caption
\myfig  {12} 0 {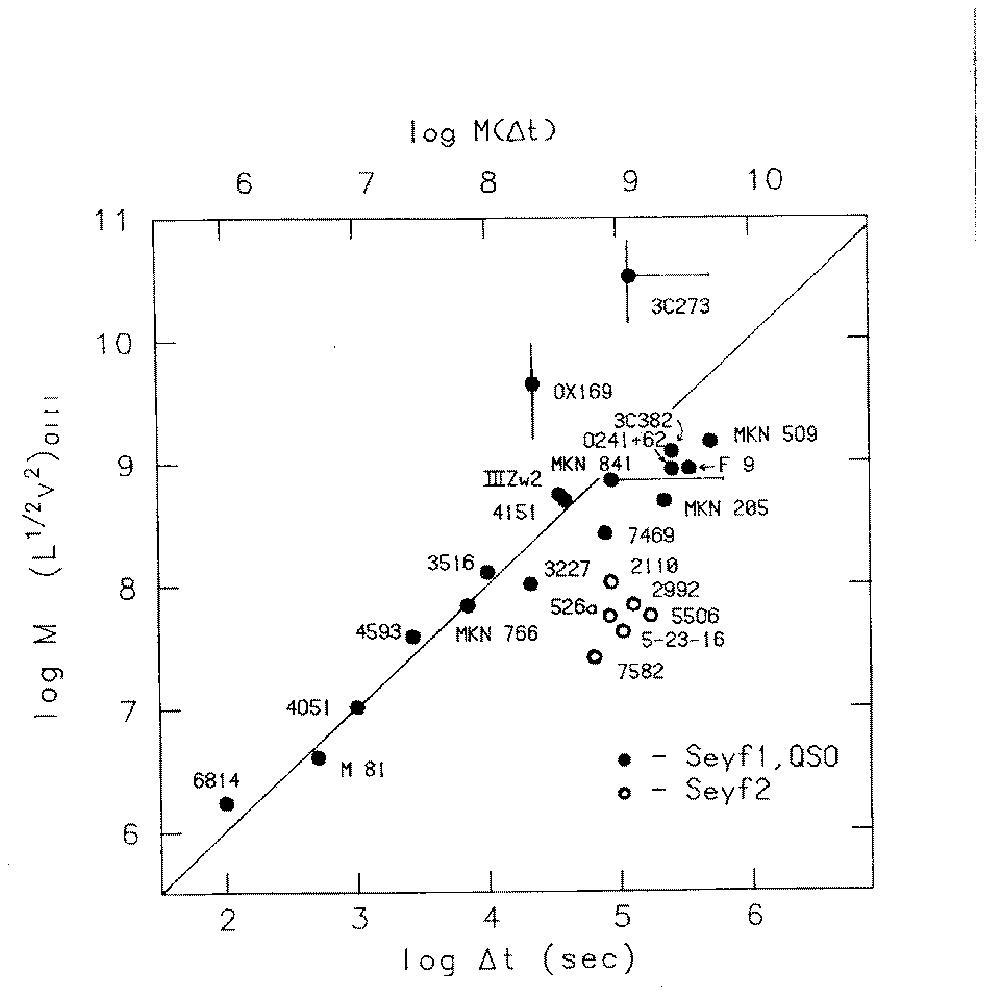}
{3. X-ray variability time versus [O~{\sc iii}]
kinematic mass estimate (Wandel and Mushotzky 1986).}
%\label {wm}\end{figure}

\section{Accretion-disk modeling}
Many authors have tried to fit \ad spectra to the observed AGN continuum, thus finding the
\ad parameters ($M, \dot M, \alpha$ and the angular momentum of the black hole)
which best fit the observed
continuum in the UV  (e.g. Wandel \& Petrosian 1988; Sun \& Malkan 1991)
or in the soft X-rays (e.g. Laor  1990).

\subsection{The \ad - \bb spectral relation}

The thin \ad spectrum is a multi-temperature \bb one, and
trying to fit the UV bump (or the soft
X-ray spectrum) is
essentially equivalent to using eq. \ref{equ:mt} in order to find the temperature-mass:
$M_8\approx 3 L_{45}^{1/2}/ T_5^2 (r/10)$ where $r=R/R_s$.
The second \ad parameter,
$\dot M$, is determined by the observed luminosity, schematically via the relation
$L=\epsilon Mc^2$.
An alternative approach relates the peak in the \ad spectrum to the thin \ad
parameters by assuming the spectrum is dominated by \bb emission near the
maximum-surface emissivity radius. This gives
\begin{equation}
E_{max}\approx 3kT_{bb}(5R_s)=10 (\dot m /M_8 )^{1/4}~{\rm eV}
\end{equation}
where $\dot m=\dot M/\dot M_{\rm Edd}$ is the \ed ratio. More sophisticated approaches,
like taking as $T_{max}$ the temperature at the inner boundary between the \bb and
the radiation-dominated disk or the temperature at which the inner disk
becomes optically thin give $ E_{max}\approx 2 (\dot m M_8)^{-0.3}$~eV.
While in the UV band the thin \ad multiple \bb spectrum may be a good
approximation, the soft X-rays may
be produced by a hotter medium due to processes other than \bb emission, such as
Comptonization, a two-temperature disk (Wandel \& Liang 1991) or hot corona
(Haardt \& Maraschi 1993; Czerny, Witt \& Zycki 1996).

%\begin{figure}\vspace {8cm}\plotone{wp.PS}
\myfig {10} 1 {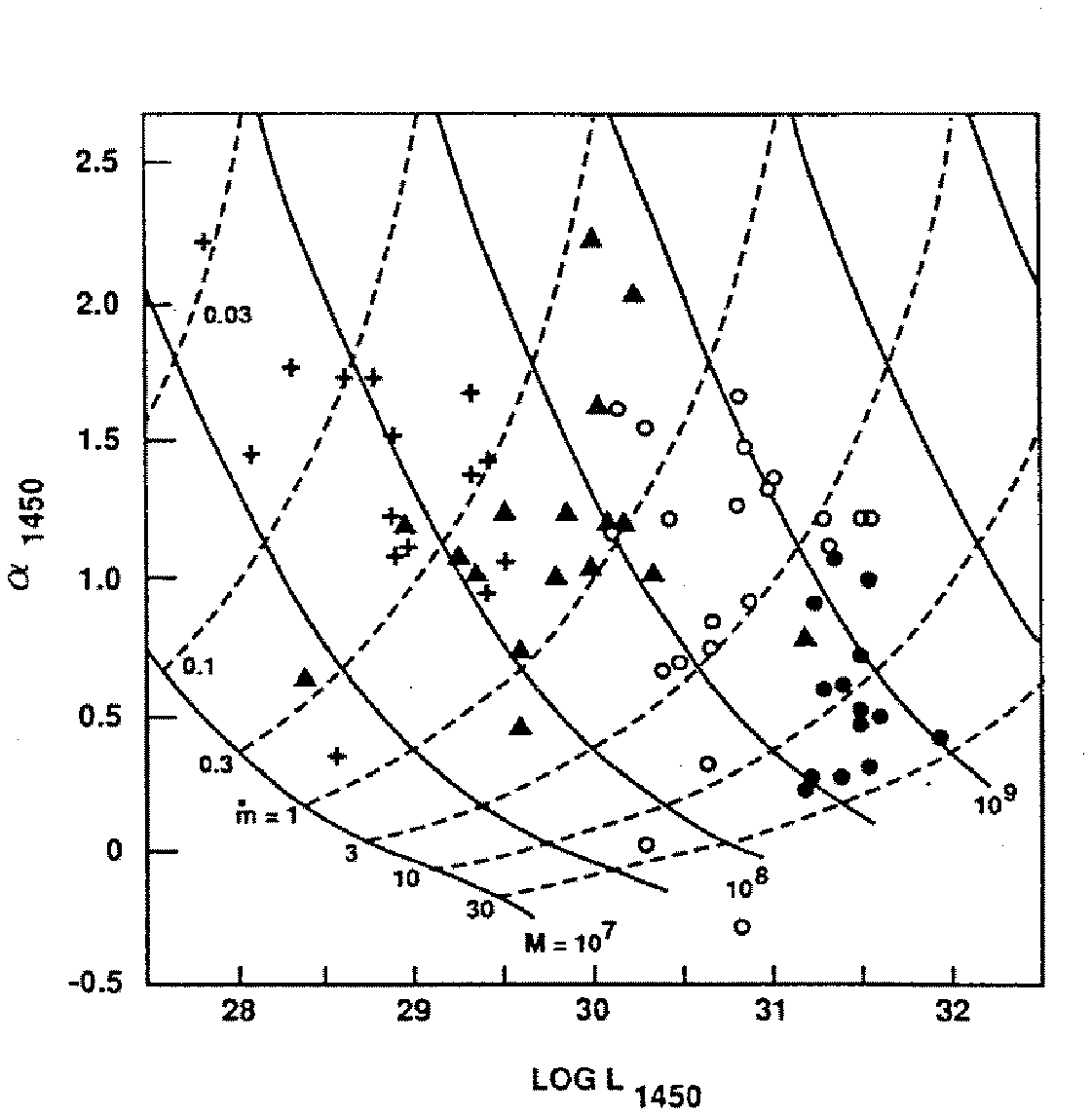}
{4. Accretion-disk \bh evolution tracks in the $\alpha -L$ plane
(Wandel \& Petrosian 1988). Crosses: Seyfert galaxies, triangles, open circles and filled circles:
low, medium and high redshift quasars, respectively. Continuous lines: constant mass;
dashed lines: constant $\dot M/M$ (note that, unlike in the text, the labeling
of the curves in the figure follows the notation
$\dot m=16.7\dot M/\dot M_{\rm Edd}$).}
%\label {wp}\end{figure}
%Evolution

\subsection{Deriving $M$ and $\dot M$ from the UV spectrum}
It is possible to perform a more refined treatment, calculating for each pair of \ad
parameters $M$ and $\dot M$ not the total luminosity and blue-bump
temperature, but the actual observables, e.g. the UV luminosity and spectral index.
Wandel \& Petrosian (1988) have calculated the \ad flux and spectral index
at 1450~\AA\ for a grid of \ad parameters
$(M, \dot M)$. Inverting the grid they have obtained
contours of constant \bh mass
 and constant \ed ratio ($\dot m$) in the $L-\alpha_{UV}$ plane (fig 4).
%~ \ref{wp}).
Plotting in this plane samples of AGN it is possible to read from the contours the
corresponding \ad parameters. Comparing several groups of AGN a systematic trend
appears: higher redshift and more luminous objects tend to have larger \bh masses
and luminosities closer to the \ed limit (see table~1). %\ref{tbl1}).
Similar results are obtained by Sun \& Malkan (1991).

\begin{table}
\caption{Grouping of AGN \ad parameters.} \label{tbl1}
\begin{center}%\scriptsize
\begin{tabular}{lccc}
AGN group & Log~$F_{1450}$ & Log $M/\ms$ & $\dot M/\dot M_{\rm Edd}$ \\
{}&{}&{}&{}\\
\tableline
%{}&{}&{}&{}\\
Seyfert galaxies & 28--29.5 &7.5--8.5 & 0.01--0.5\\
Low $z$ quasars & 29--30.5 &8--9 & 0.02--0.1\\
Medium $z$ quasars & 30--31.5 & 8--9.5 & 0.1--0.5\\
High $z$ quasars & 31--32 & 9--9.5 & 0.03--2\\
\tableline
\end{tabular}
\end{center}
\end{table}

%Summary table
\begin{table}
\caption{$L/M$ ratios found by various methods.} \label{tbl2}
\begin{center}%\scriptsize
\begin{tabular}{lll}
Method & $L/L_{\rm Edd}$ & Reference \\
%{}&{}&{}\\
\tableline
{}&{}&{}\\
{\bf Kinematic} &{}&{}\\
Emission Volume (H$\beta$)&0.003--0.03 &Wandel \& Yahil (1995)\\
Ionization parameter (H$\beta$)&0.01--0.1 &Joly et~al. (1995)\\
%\ip (CIV)&0.01--0.1 &Blumenthal and Wandel 1998\\
IP + Reverberation(H$\beta$)&0.01--1 & Wandel (1997)\\
%{}&{}&{}\\
\tableline
{}&{}&{}\\
{\bf X-ray variability}&{}&{}\\
X-rays & 0.003--0.1 & Barr \& Mushotzky (1986) \\
X-rays+OIII photoionization & 0.003--0.1 & Wandel \& Mushotzky (1986) \\
Soft X-rays, NLS1 & 0.001--0.1 & Wandel \& Boller (1998)\\
%{}&{}&{}\\
\tableline
{}&{}&{}\\
{\bf Accretion-disk fitting}&{}&{}\\
Modified BB disk & 0.001--2& Wandel \& Petrosian (1986)\\
Kerr \bh disk&0.01--2& Sun \& Malkan (1991)\\
GR effects& 0.003--0.3& Laor (1990), Laor (1993)\\
Disk+Corona& 0.001--0.3& Czerny, Witt \& Zycki (1996)\\
%{}&{}&{}\\
\tableline
\end{tabular}
\end{center}
\end{table}

\section{Summary}
The $L/M$ ratios found by various methods are summarized in table 2. %\ref{tbl2}.
All \el kinematic methods give a relatively narrow $L/M$ range over a large range in
luminosity, suggesting a universal $L/M$ relation in AGN. However, this may be
related to the fact that the samples with
available data are mostly low redshift, relatively low-luminosity objects.
The two other methods --- X-ray variability and \ad fitting of the UV bump ---
suggest a trend of the \ed ratio increasing with luminosity.

\begin{question}{Brad Peterson}
I believe that a fairly reliable virial mass can be obtained from reverberation
mapping data alone. The continuum emission-line cross-correlation function gives
a centroid that measures the responsivity-weighted radius of the BLR. The
spectra can be combined to form mean and rms  spectra. The rms spectrum
identifies the variable part of the emission line, and the line width used for
virial estimates should be that measured from the rms spectrum. This way both
the characteristic line width and scale size refer to the same gas, that which
is varying  at the time of the experiment.
\end{question}
\begin{answer}{Amri Wandel} I fully agree. This method actually gives ``the best of
both worlds'': a direct measurement of the distance along with the appropriately
weighted velocity. It is interesting to compare the rms reverberation  and the
ionization-parameter methods, in order to calibrate the latter, which is  used
more readily and also for objects without variability data.
\end{answer}
\begin{question}{Michael Corbin}
I wish to comment that  there exists a strong correlation between the
asymmetry of the C~{\sc iv}~$\lambda$1549 and the H$\beta$ profiles and the luminosity of the soft
X-ray and near UV continuum, such that as luminosity increases, the profiles
become more red asymmetric. I have been able to model these redward asymmetries
as the effect of gravitational redshift, which is proportional to the mass of
the black hole. This correlation is therefore consistent with  a universal \bh
mass-luminosity relationship.
\end{question}
\begin{question}{Martin Gaskell} It is interesting that the \ad method shows $M$
tending to $M_{\rm Edd}$ when we go to high luminosities. From reverberation mapping
(Koratkar \& Gaskell 1991, ApJL) we found 3C273 near $M_{\rm Edd}$. Prab Gondhalekar
has found this to be generally true for high luminosity quasars. However, from
our analyses of 3C273, I think that beaming of the continuum is causing an
underestimate of the radius in reverberation mapping.
\end{question}
\begin{answer}{Amri Wandel} The trend of the $L/M$ ratio to be larger for more luminous
objects appears to be associated only with the \ad fitting method of mass
estimation. When the mass is estimated by the \ip method, this trend is much
weaker or even consistent with $M\propto L$ (hence $L/M\sim $ const).  The
correlation between $L/M$ and the luminosity in the \ad method may therefore
reflect a bias related to the thin \ad equations or the UV-X-ray continuum of
AGN.
\end{answer}
\begin{question}{Jean Clavel} Observationally, how does one define a characteristic
``fastest'' X-ray variability? The power-density has a slope close to unity (e.g.
is self-invariant), so that there is no characteristic time scale and the
probability of observing an event of  $\Delta L$ within a time $\Delta t$
increases with the length of the observation.
\end{question}
\begin{answer}{Amri Wandel} The variability time scale has been determined only from
large events (change of the flux  by a factor of 1.5 or more), and the
characteristic time has then been taken as the e-folding time of that change.
\end{answer}

\end{document}